\documentclass[%
 aip,
 jcp,%
 amsmath,amssymb,
 reprint,%
]{revtex4-2}

\usepackage{graphicx}
\usepackage{dcolumn}
\usepackage{bm}
\usepackage[version=4]{mhchem} 
\usepackage[T1]{fontenc}
\usepackage{graphicx}
\usepackage{bm}
\usepackage{hyperref}
\hypersetup{
    colorlinks=true,
    linkcolor=blue,
    filecolor=blue,
    urlcolor=blue,
   citecolor=blue,
}
\usepackage{siunitx}
\usepackage{enumitem}
\usepackage{amssymb}
\usepackage{xcolor}
\usepackage{soul}

\begin{document}


\title{Phase Transitions Induced by Resonant Light: \\A Phenomenological Approach}

\author{A. Kudlis}
\email{akudlis@hi.is}
\affiliation{Science Institute, University of Iceland, Dunhagi-3, IS-107 Reykjavik, Iceland}
\author{L. S. Ricco}
\affiliation{Science Institute, University of Iceland, Dunhagi-3, IS-107 Reykjavik, Iceland}
\author{H. Sigur{\dh}sson}
\affiliation{Science Institute, University of Iceland, Dunhagi-3, IS-107 Reykjavik, Iceland}
\affiliation{Institute of Experimental Physics, Faculty of Physics, University of Warsaw,\\ ul.~Pasteura 5, PL-02-093 Warsaw, Poland}
\author{I. A. Shelykh}
\affiliation{Science Institute, University of Iceland, Dunhagi-3, IS-107 Reykjavik, Iceland}

\date{\today}

\begin{abstract}
We present a phenomenological framework to describe a subclass of light-induced phase transitions (LIPTs) in condensed matter systems, specifically those mediated by the resonant generation of excitons. Our approach extends the classical Landau theory by introducing dynamic coupling between the system's order parameter and complex excitonic fields, along with Langevin-type forces that drive the system toward states of minimal free energy. The model is applied in the context of all-optical resonant magnetization switching in two-dimensional magnetic materials, particularly reproducing the experimental findings for reverse magnetization by all-optical means for a monolayer CrI$_3$.
Our phenomenological model can be applied to other systems characterized by an order parameter and excitonic fields created through resonant light, offering versatility and potential to guide future experimental and theoretical studies in LIPT phenomena. 
\end{abstract}

\maketitle

\section{Introduction}

The concept of a phase transition is among the foundational tenets of condensed matter physics. A state of a macroscopic system can undergo an abrupt change when external control parameters, such as temperature, pressure, or magnetic fields, reach certain critical values~\cite{stanley1971phase,10.1093/acprof:oso/9780199227198.001.0001}. Understanding the mechanisms of phase transitions is crucial for elucidating the behavior of the materials at both atomic and macroscopic scales, as well as for exploring new technological applications. 

The standard phenomenological formalism for the description of phase transitions is provided by the well-established Landau theory~\cite{Landau1937I,Landau1937II}, which operates with the system's free energy expressed as a function of its order parameter expanded in its powers, with coefficients of expansion depending on temperature and, possibly, other parameters defining the state of the system. Further development of the theory allowed to include the phenomena of critical fluctuations and non-equilibrium effects. Moreover, new powerful theoretical methods such as renormalization group \cite{10.1093/acprof:oso/9780199227198.001.0001,Ma1980,Patashinsky1982,Cardy1996,Vasiliev1998,PhysRevB.4.3174,PhysRevB.4.3184}, and conformal bootstrap~\cite{Polyakov1970,Polyakov1974,Belavin1984,Rattazzi2008,Poland2016,RevModPhys.91.015002} were developed. 

In recent years, employing light as an external control parameter has opened new avenues for exploring the system's phase transitions. Laser pulses can modify the energy landscape of the materials on femtosecond timescales, thus driving ultra-fast phase transitions with mechanisms that differ fundamentally from those driven solely by e.g. thermal effects~\cite{Nasu2004}. Usually, a combination of various effects, including thermal, non-thermal, excitonic, and other processes, plays a role in so-called light-induced phase transitions (LIPTs). Some few examples of LIPTs in strongly correlated systems are the sub-picosecond insulator-to-metal transitions in \ce{VO2} via pump-probe experiments~\cite{cavalleri2001femtosecond,wegkamp2014instantaneous,johnson2023ultrafast}, reshaping of charge density wave (CDW) phase in 1T-\ce{TaS2}~\cite{hellmann2010ultrafast,schmitt2008transient}, transition to photoinduced high-spin phase in [Fe(2-pic)]$_3$Cl$_2$~\cite{Ogawa2000,Tayagaki2001,Huai2002,Nasu2004}, ferroelectric transition in strontium titanate~\cite{Nasu2004}, all-optical phase switching in ferromagnets and multiferroics~\cite{kimel2005ultrafast,stanciu2007all,Dabrowski2022,Zhang2022,PhysRevB.104.L020412,Kazemi2024,PhysRevB.108.094421,beaurepaire1996ultrafast,kirilyuk2010ultrafast}, photoinduced transient superconductivity~\cite{fausti2011light,mitrano2016possible}, photoinduced structural changes in two-dimensional (2D) perovskites~\cite{Kandada2020ExcitonPolarons, Tao_NatComm_12_2021, Zhang_NatPhys_2023, Thouin_NatMat_2019} and isomerization~\cite{doi:10.1021/acs.chemmater.6b03460,ichikawa2011transient,wall2011ultrafast,collet2003light,kirschner2019}.

To explain and reproduce the findings in the field of LIPTs, a variety of corresponding theoretical models were also developed, as discussed in Refs.~\cite{Nasu2004,giannetti2016ultrafast,freericks2006theoretical,aoki2014nonequilibrium}. However, most existing approaches focus on specific systems or rely on detailed microscopic models that may not be readily generalizable. A unified model, grounded in the well-established framework of Landau theory but extended to include non-thermal light-driven effects, offers a promising route to bridge these gaps. Such a model would allow for direct comparison between different systems, facilitate the systematic extraction of key parameters, and provide predictive insights into material behavior under optical excitation.

In this paper, we take a step further to build a generalized model for LIPTs by presenting a phenomenological framework specifically addressing a class of phase transitions associated with two distinct excitonic modes excited by resonant light. Our model is based on the classical Landau theory of phase transitions but incorporates key elements to describe the proposed class of resonant LIPTs, including the interplay between Langevin-type dynamics, complex excitonic fields, and temperature. To show our model's capability, we explore the representative case of all-optical magnetization switching in van der Waals ferromagnets such as \ce{CrI3}, where a resonant pulse of photons carrying $\sigma^\pm$ circular polarization selectively excite spin $\uparrow, \downarrow $ exciton states via angular momentum transfer and changes the order parameter (magnetization) from “up’’ to “down’’ with negligible heating. We also discuss the potential applications of our framework to other systems non-related to magnetic ordering, such as optically induced structural changes in 2D metal halide perovskites mediated by excitons. Our developed Landau-type model allows to generically describe ultrafast all-optical switching between macroscopic states that are governed by the same underlying mechanism: the resonant enhancement of one phase associated with the creation of an excitonic reservoir that lowers its corresponding free-energy minimum.   

The paper is organized as follows. We first introduce the general model and analyze its qualitative behavior under both continuous and pulsed resonant optical excitations. We then apply the model to the specific scenario of all-optical magnetic switching in \ce{CrI3}.  Finally, we qualitatively discuss the possibility of applying our framework to other systems such as Ruddlesden-Popper lead-halide
perovskites and conclude with a summary of the main results.

\section{The Model}

\begin{figure}[t]
    \centering
    \includegraphics[width=1\linewidth]{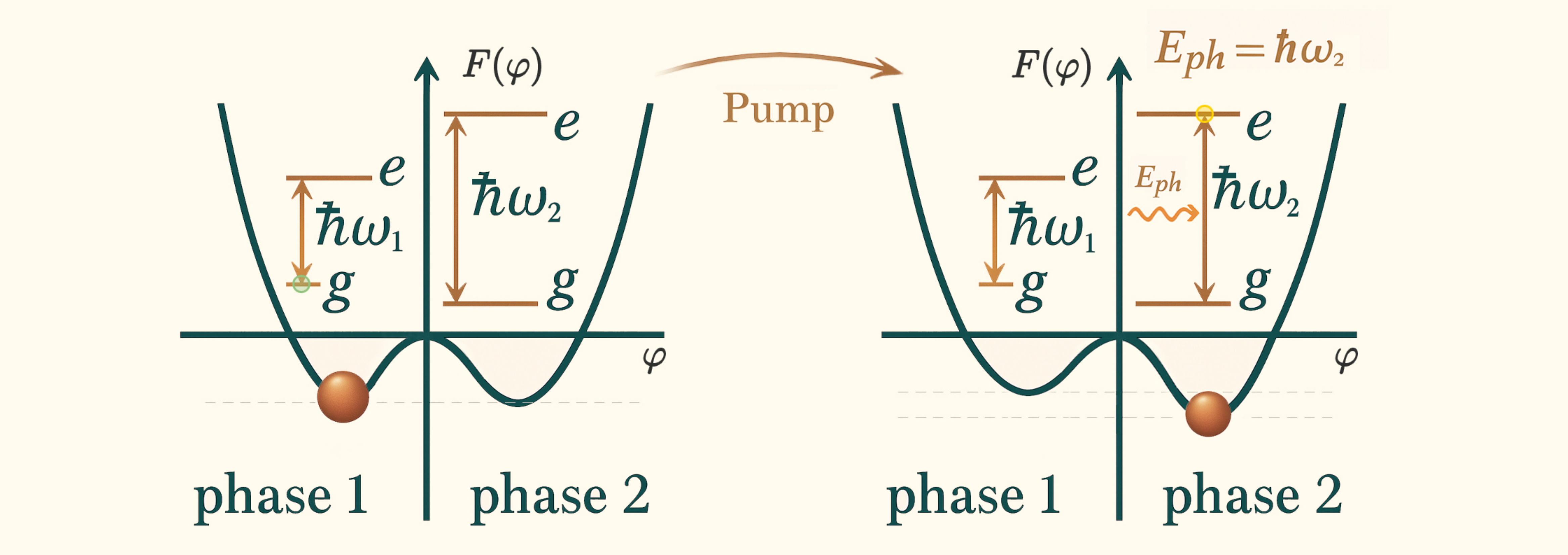}
    \caption{Conceptual illustration of the resonant-exciton mechanism that underlies our Landau-type model for light-induced phase switching. The red dot shows schematically in which phase the system resides. Left panel: Free energy $F(\varphi)$ of the system as a function of the order parameter $\varphi$, characterizing the phase transition in the absence of external optical excitation. Two stable phases correspond to the two minima of the double-well potential with both ground (g) and excited (e) states. They are characterized by different energies of optical excitonic transitions $\hbar\omega_1\neq\hbar\omega_2$. When the system resides in phase 1 (left well), the corresponding excitonic population is zero. Right panel: The system's free energy profile is reshaped under an external optical pump resonant with the excitonic transition in phase 2, $E_{ph}=\hbar\omega_2$. The enhancing probability to create excitons in phase 2, but not in phase 1, drives the system into a new phase by lowering the corresponding free energy minimum, which is conditioned by the cubic coupling $\gamma\!\propto\!|\psi_1|^{2}-|\psi_2|^{2}$ that tilts the landscape. The additional Langevin noise allows the switching without raising the lattice above its critical temperature.}   \label{fig:LIPTscheme_2DIsing}
\end{figure}

We consider a system characterized by two distinct ordered phases below the classical phase transition driven by temperature $T$. Following the standard practice in the Landau theory of phase transitions, we also postulate the existence of an order parameter $\varphi$ that defines the free energy profile. This profile is expected to have two minima, corresponding to two stable phases. For example, in a 2D Ising ferromagnetic system, such as a CrI$_3$ monolayer (see the following section), the relevant order parameter is magnetization, which preferentially aligns either "up" or "down" with respect to the layer interface. 

According to the phenomenological Landau model, the free energy is given by
\begin{align} 
F(\varphi, T) = \alpha(T) \varphi^2 + \beta \varphi^4, \label{eq:Landau_nocubic} 
\end{align}
where $\beta>0$ and $\alpha(T) = \alpha_0 (T - T_0)$, with $\alpha_0>0$ being a material-specific constant, and $T$ and $T_0$ being the system's temperature and phase transition temperature, respectively. We suppose that the system's two phases differ in their electronic band structure; in particular, the corresponding energies of optical excitonic transitions are substantially different in two phases $\hbar\omega_1\neq\hbar\omega_2$, as shown in the left panel of Fig.~\ref{fig:LIPTscheme_2DIsing}. The free energy described by Eq.~\eqref{eq:Landau_nocubic} clearly preserves the $\mathbb{Z}_2$ symmetry and both the excitonic phases are equivalent.

Let us start by supposing that the system resides in phase 1. If an optical pump is absent, there are no excitons in the system for $k_BT\ll\hbar\omega_{1,2}$, where $k_B$ is the Boltzmann constant. Thermal fluctuations can in principle drive the system spontaneously from phase 1 to phase 2, but if the energy barrier separating the two phases is much higher than the thermal energy, such transitions are highly improbable. 

Now, let us consider that the system is pumped by an external laser, such that the energy of the photons is put in resonance with the excitonic transition energy in phase 2, $E_{ph}=\hbar\omega_2$. This optical pump will produce two effects. The first one is obvious: non-resonant background absorption will heat the sample, so that the value of the potential barrier separating the phases becomes lowered, and the role of the thermal fluctuations will increase. For very high pumps, when the sample is heated above the phase transition temperature, $T>T_0$, this barrier will simply disappear, merging the two energy minima into a single one, and the system will undergo into a non-ordered phase. The second effect induced by the resonant pump is the mechanism explored in our work. In fact, since $E_{ph}\neq\hbar\omega_1$, the creation of excitons in phase 1 is inefficient. However, it becomes possible in phase 2, where the resonant condition $E_{ph}=\hbar\omega_2$ is met. This will drive the system into phase 2, lowering the corresponding free-energy minimum, as shown in the right panel of Fig.~\ref{fig:LIPTscheme_2DIsing}. This happens because the absorption of the coherent light results in the creation of an excitonic reservoir, where the processes of non-radiative excitonic recombination lead to the transformation of the energy from photons into heat, increasing the system's entropy and decreasing its free energy. Therefore, the corresponding phase transition mechanism does not rely solely on heating; rather, it is triggered by the resonant enhancement of excitons associated with one particular phase, thereby breaking the symmetry of the potential wells and rendering the local minima of the free energy. 

In this sense, our model intends to generically describe transitions driven by resonant light that unbalance excitonic populations between two phases, thereby reshaping the free energy landscape and altering the exciton-related order parameter $\varphi$. For instance, $\varphi$ may represent the magnetization in \ce{CrI3} layers, as explored in detail here, or the lattice distortion amplitude in 2D lead-halide perovskites~\cite{Tao_NatComm_12_2021}, where the transition involves resonantly induced switching between free excitons and self-trapped exciton polarons. 

The minimal way to capture the effect described above within Landau’s framework is to introduce a cubic term into the free energy, 
\begin{align} 
F(\varphi, T) = \alpha(T) \varphi^2 + \gamma \varphi^3 +\beta \varphi^4, \label{eq:Landau} 
\end{align}
which explicitly breaks the $\mathbb{Z}_2$ symmetry in relation to $\varphi$, and it is associated with the imbalance between two excitonic phases. To catch this imbalance, the parameter $\gamma$  depends on the difference between excitonic populations corresponding to phases 1 and 2, i.e.:
\begin{align}
    \gamma= \gamma_0 (|\psi_1|^2-|\psi_2|^2),\label{eq:gamma}
\end{align}
where $\gamma_0$ is a positive constant, while $\psi_j$ corresponds to excitonic fields associated with two different phases ($j=1,2$), with corresponding excitonic occupancies $n_j=|\psi_j|^2$ that are modulated by the external optical pump. In this way, the set of Eqs.\eqref{eq:Landau}-\eqref{eq:gamma} allows to generically describe a $\mathbb{Z}_2$ symmetry breaking dynamically mediated by the optically selected excitonic phases via external pump, once the sign of $\gamma$ defines which phase will have the lowest free energy. It can be extended to any system in which such an inversion symmetry breaking is triggered by two distinct excitonic phases controlled externally and capable of governing the exciton population dynamics without the requirement of knowing the system´s microscopic details.

The inclusion of an exciton excitation imbalance through the cubic term $\gamma\varphi^3$ is motivated by both symmetry and physical considerations. As discussed previously, this term explicitly breaks the inversion ($\mathbb{Z}_2$) symmetry of the order parameter in Eq.~\eqref{eq:Landau} and leads to a first-order phase transition driven by the exciton population imbalance encoded in $\gamma$. In contrast, although a linear term $\gamma\varphi$ would also break inversion symmetry, it acts as an effective external field conjugate to the order parameter, biasing one phase at all temperatures and smoothing out the transition into a crossover, which is inconsistent with a first-order phase transition we intend to describe here. A similar line of reasoning is employed for explaining the effect of photoinduced phase separation in multi-component lead-halide perovskites~\cite{Limmer_JCP_10.1063/1.5144291(2020),Draguta_NatComm_2017}, where a linear change in the local composition cannot account for the emergence of phase separation. Instead, a highly nonlinear response is required to set the local composition at a given temperature, corresponding to a bistability in the free energy landscape.
Moreover, the assumption of a quadratic term $\gamma\varphi^2$ would renormalize the existing $\alpha(T)\varphi^2$ on Eq.~\eqref{eq:Landau} and does not introduce any qualitatively new physics to describe a phase-transition mediated by the excitonic population imbalance. We should stress again, for the avoidance of doubt, that while such lower-order terms may well arise microscopically -- with the linear term biasing the relative stability (conjugate-field effect) and the quadratic term primarily renormalizing $\alpha(T)$ and the effective proximity to $T_c$ -- in our minimal model we omit them to isolate the odd-in-$\varphi$ barrier-tilting mechanism responsible for first-order, threshold-like switching; their inclusion would chiefly shift thresholds and detuning windows without altering the central phenomenology.

The free energy shape given by $F$ defines the dynamic equation for the order parameter $\varphi$ according to:
\begin{equation}
 \frac{d\varphi}{dt} = -\frac{1}{\tau_\varphi} \dfrac{\partial (F/F_0)}{\partial \varphi} + \eta(t). \label{eq:EOM_phi}  
\end{equation}
The first term in the relation above describes the relaxation to a minimal energy state characterized by relaxation time $\tau_\varphi$, where $F_0$ is a characteristic free energy scale, taken as $\alpha_0F_0=T_0$. The second term accounts for the role of thermal fluctuations, which are important for systems undergoing phase transitions, especially near critical points. It can be chosen as a stochastic Langevin force with statistical properties characteristic of Gaussian white noise as follows:
\begin{align} &\langle \eta(t) \rangle = 0, \\ 
&\langle \eta(t) \eta(t') \rangle = 2 D(t) \delta(t - t'),  \\
& D(t) = \frac{k_B T(t)}{F_0\tau_\varphi},\label{eqn:noise}
\end{align}

\noindent In equilibrium ($P=0$), the choice $D(t)=k_B T(t)/(F_0\tau_\varphi)$ implements the fluctuation–dissipation relation for the relaxational dynamics in Eq.~\eqref{eq:EOM_phi}, ensuring detailed balance. Out of equilibrium ($P\neq 0$), the same expression defines a time-dependent noise strength slaved to $T(t)$, thereby providing a back-action channel $|\psi_j|^2\!\to\!T(t)\!\to\!D(t)\!\to\!\varphi$ in addition to the deterministic odd-in-$\varphi$ contribution to the free energy.

The evolution of the complex excitonic fields $\psi_j(t)$ is governed by an external optical pump $P(t,\omega)$ as follows:
\begin{equation}
i\hbar\frac{d\psi_j}{dt} = \epsilon_j \psi_j(t) +a| \psi_j(t)|^2\psi_j(t) + S_j(\varphi) P(t,\omega), \label{eq:psi_dynamics} 
\end{equation}
with $j=1,2$, where $\epsilon_j=\hbar\omega_j-i\gamma_j$, with $\gamma_j$ being excitonic broadenings directly related to their decay rates and $\omega_j$ the excitonic frequency. The parameter $a$ quantifies the strength of the nonlinear exciton-exciton interaction, which arises from Coulomb repulsion between excitons, such as in the case of Wannier excitons. This term captures the density-dependent energy shift of the excitonic resonance induced by local exciton populations. The resulting nonlinearity plays a central role in describing energy blueshifts, optical bistability, and nonlinear density-dependent responses. In addition, the nonlinear term is responsible for preventing unbounded growth of $|\psi_j|$ under pumping, ensuring a saturation mechanism to the excitonic dynamics, and should be kept even for small $a$. 

Let us note, that in the weakly coherent (fast-dephasing) regime, where $\gamma_j^{-1}\!\ll$ the pulse duration and the pump-induced coherent coupling is small, $|S_j(\varphi)P(t,\omega)|/\hbar \ll \gamma_j$, the optical phase can be adiabatically eliminated. Then the driven-field equation yields $\psi_j(t)\!\approx\!S_j(\varphi)P(t,\omega)\big/\!\big(\hbar(\omega-\omega_j)+i\gamma_j-a|\psi_j|^2\big)$, and the exciton populations $n_j=|\psi_j|^2$ obey
\begin{align}
\dot n_j =\! -\Gamma_j n_j\! +\! \kappa_j\, S_j(\varphi)^2|P(t)|^2L_j\!\big(\omega\!-\!\omega_j\!-\!\delta\omega_j(n_j)\big),
\end{align}
with $L_j(\Delta)=\gamma_j^2/(\Delta^2+\gamma_j^2)$ and a weak density-induced shift $\delta\omega_j\propto a n_j/\hbar$. Here we define the population decay rate as $\Gamma_j = 2\gamma_j/\hbar$ and absorb the remaining coupling constants into $\kappa_j$; with the present normalization of $L_j$, one may set $\kappa_j = 2/(\hbar\,\gamma_j)$.

In Eq.~(\ref{eq:psi_dynamics}), we introduce a shape-selective factor $S_j(\varphi)$, defined as
\begin{align}
S_j(\varphi)=\frac{1}{1+\displaystyle\bigl[(\varphi-\varphi_j^{(0)})/\Delta\varphi\bigr]^2},
\label{eq:psi_dynamics_sel_S}
\end{align}
where $\varphi_j^{(0)}$ is the position of the free energy minimum corresponding to phase $j$ and $\Delta\varphi$ is the width which controls the \textit{blurriness} of the boundary between phases. The factor $S_j(\varphi)$ is essential because each ordered phase possesses its own excitonic transition, so optical absorption, say, at $\hbar\omega_2$ is negligible while the order parameter remains near $\varphi^{(0)}_1$ and vice-versa.  By modulating the pump term with $S_j(\varphi)$, we ensure that excitons $\psi_2$ can be created only after the structure has begun to adopt the geometry of phase $2$. Otherwise, a phase-independent pump would instantaneously generate $|\psi_2|^2$ inside phase $1$, due to a large negative contribution to $\gamma=\gamma_0(|\psi_1|^2-|\psi_2|^2)$, and artificially tilt the Landau potential before any real structural change occurs.  The shape-selective factor term therefore eliminates this unphysical \textit{teleportation} of excitons, establishing self-consistent feedback between optical excitation and the evolving order parameter, and yields the experimentally observed threshold and hysteresis in resonant all-optical switching phenomena. Quantitatively, the results are robust with respect to the choice of $\Delta\varphi$: varying it by an order of magnitude around our nominal value preserves all qualitative features (existence of a fluence threshold, helicity selectivity, and strong detuning dependence), affecting mainly the threshold position. In the hard-selection limit $\Delta\varphi\!\to\!0$, switching still occurs but requires larger stochastic excursions and/or fluence to nucleate entry into the narrow overlap region where $S_j(\varphi)$ becomes effective; conversely, for larger $\Delta\varphi$ the exciton reservoir develops earlier along the trajectory and the threshold decreases moderately. Physically, $\Delta\varphi$ serves as a proxy for microscopic inhomogeneity and finite domain-wall width (and weak off-resonant channels), and its impact is most pronounced for ultrashort pulses where reservoir build-up is time-limited.

Considering that the complex field $\psi_j$ describing the excitons is a canonical variable, the relation $i\hbar(d\psi_j/dt) = (\partial \mathcal{H}/\partial\psi^{*}_j)$ holds. Then, the Hamiltonian that corresponds to the equation of motion describing the exciton dynamics given by Eq.~\eqref{eq:EOM_phi} reads: 
\begin{equation}
    \mathcal{H} = \epsilon_j |\psi_j|^2 + \frac{a}{2}|\psi_j|^4 + S_j(\varphi)P(t,\omega)[\psi_j(t) + \psi^{*}_j(t)]. \label{eq:Hamiltonian}
\end{equation}
In principle, an extra term in the Hamiltonian above would couple the excitonic field to the order parameter $\mathcal{H}_{\gamma}\propto \gamma(|\psi_1|^2- |\psi_2|^2)\varphi^3$, to ensure reciprocity with Eqs.~\eqref{eq:Landau}-\eqref{eq:gamma}, where $\psi_j$ and $\varphi$ are explicitly coupled via $\gamma$. This would also introduce the extra term $\delta_j\gamma_0\psi_j\varphi^3$ in
Eq.~\eqref{eq:psi_dynamics} governing the exciton dynamics, with $\delta_j=+1,-1$ for $j=1,2$, describing the feedback mechanism from $\varphi$ to $\psi_j$. However, notice that our focus is to describe experimental situations wherein the order parameter $\varphi$ and the associated phase transition primarily respond to the excitonic population imbalance, mostly governed by the external pump $S_j(\varphi)P(\omega,t)$, as will be explored later for the case of all-optical magnetization switching in \ce{CrI3}. Therefore, the influence of $\varphi$ on the exciton dynamics can be safely neglected. In addition, we note that the inclusion of such a back-action term on the exciton dynamics would lead only to a small renormalization of the excitonic energies, with $\tilde{\epsilon}_j=\epsilon_j + \delta_j\gamma_0\varphi^3$, with $\varphi^3 \ll 1$ near the phase transition.

The final component in the proposed model describes the temporal evolution of the temperature as follows:
\begin{equation} 
\frac{dT}{dt} = -\frac{1}{\tau_T} (T(t)-T_r) + a_T |P|^2 + \sum_j b_j |\psi_j|^2, \label{eq:Temperature}
\end{equation}
where $\tau_T$ is the constant governing the temperature decay in the absence of external sources, $T_r$ is the temperature of the environment.
In Eq.~\eqref{eq:Temperature}, the term $a_T|P|^2$ accounts for background/off-resonant absorption that does not proceed via the resonant exciton reservoirs (e.g., substrate absorption, phonon-assisted tails, or higher-lying bands weakly addressed by the pump). In contrast, the contribution $\sum_j b_j|\psi_j|^2$ captures heating due to nonradiative relaxation of the optically generated excitons and subsequent energy transfer to the lattice. 
At a coarse-grained level, one may parameterize $b_j$ as proportional to the nonradiative fraction of the decay channel and its rate. 
The stochastic force in Eq.~\eqref{eq:EOM_phi} is taken Gaussian and $\delta$-correlated, corresponding to the large-bath (Markovian) limit in which the lattice/phonon environment acts as a thermal reservoir. Regarding the thermodynamic consistency we commented it out after Eq.~\eqref{eqn:noise}.

Together with the Landau free-energy [Eq.~\eqref{eq:Landau}], the set of coupled Eqs.~\eqref{eq:EOM_phi},~\eqref{eq:psi_dynamics} and~\eqref{eq:Temperature} forms the phenomenological model hereby proposed to describe LIPTs for the case of excitonic transitions created by resonant pump. 

\section{Results and Discussion}

\subsection{Qualitative Analysis}\label{sec:quali_analysis}

\begin{figure}[b!]
\centering
\includegraphics[width=0.99\linewidth]{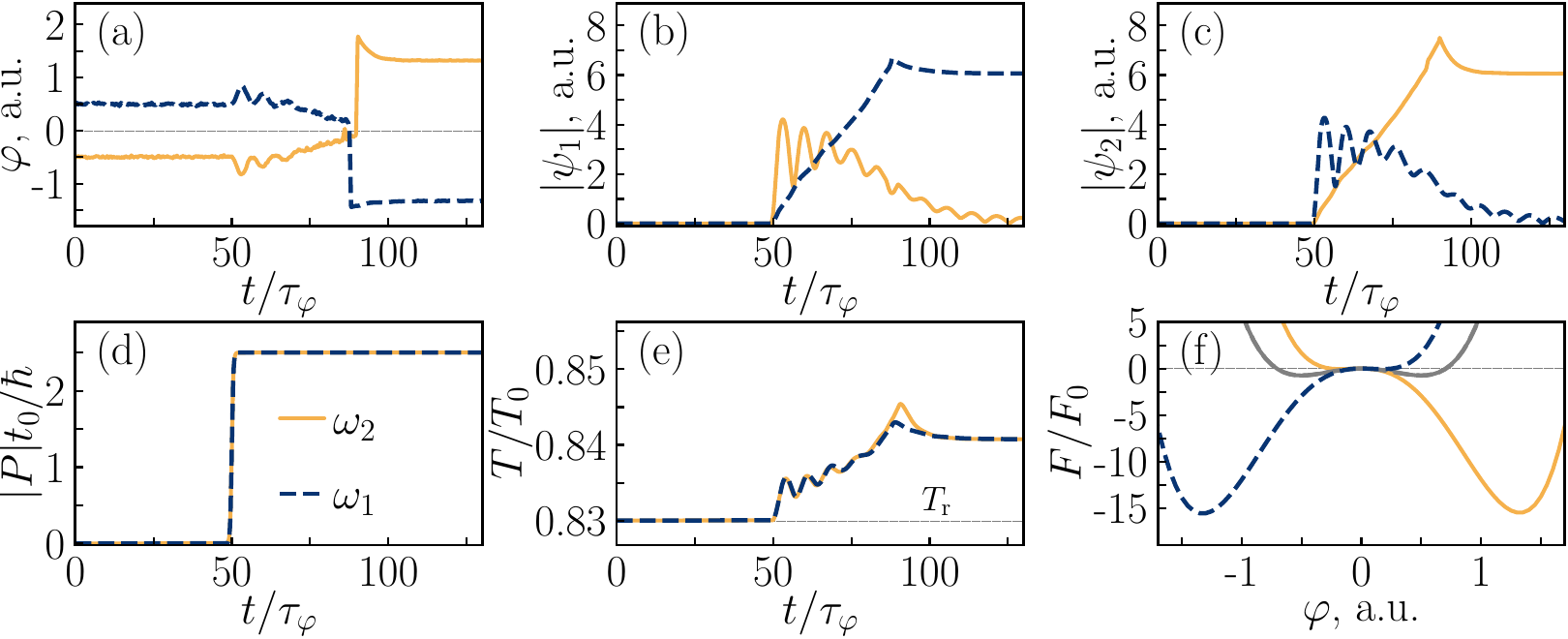}
\caption{Time-domain response of the coupled order-parameter / exciton / temperature model deep inside the ordered regime, $T_r<T_0$.  
Deep blue dashed curves correspond to a pump tuned exactly to the $\omega_1$ exciton of \emph{Phase 1}; Yellow solid curves are obtained for the different frequency, resonant with the $\omega_2$ exciton of \emph{Phase 2}.  
(a)~Dependence of the order parameter $\varphi(t)$ on time. The resonant drive that is tuned to a frequency corresponding to a different phase from that in which the system is at the initial moment in time, first, reduces the metastable barrier and, aided by noise, flips $\varphi$ from the initial minimum to the opposite well.
(b,c)~Corresponding populations of the two excitonic reservoirs $|\psi_{1,2}|^{2}$. Only the exciton that is in resonance with the pump accumulates strongly, illustrating the phase-selective nature of optical absorption built into Eq.~\eqref{eq:psi_dynamics_sel_S}.  
(d)~Temporal envelope of the square pump pulse ($t_0/\tau_\varphi\!=\!50$, fixed width) applied at the two different photon energies.  
(e)~Transient lattice temperature.  In both cases heating remains modest ($T/T_0<0.86$), confirming that reversal is non-thermal and driven by the cubic exciton term rather than by crossing $T_0$.  
(f)~Landau free-energy density $F(\varphi)$ at $t=0$ (grey, identical for both runs) and at $t=90\tau_\varphi$ after the pulse has acted. In the yellow trace the cubic contribution has tilted the landscape, making the right-hand minimum the global energy minimum and leaving the system in \emph{Phase 2}; in the deep blue trace the situation is opposite.  Together the six panels demonstrate deterministic, excitonic-state-selective phase switching governed by resonant exciton pumping in the low-temperature limit.}
\label{fig:2}
\end{figure}

In what follows, we qualitatively examine the behavior of the proposed model under simple parameter conditions to determine whether it successfully captures the essential features we originally intended: a phase transition induced by light which is responsible for resonantly creating excitonic populations in one phase, lower its corresponding free energy minimum and for switch the exciton-mediated order parameter. Figure~\ref{fig:2} shows a representative example of the behavior predicted by our framework. We plot two curves corresponding to optical pumping at different frequencies, each resonant with one of the two excitonic phases. In these two cases, we also choose different initial values of the order parameter $\varphi(t=0)$ [see panel (a)], thereby demonstrating switching from phase 1 to phase 2 (and vice versa) depending on the chosen initial state. The results are relatively insensitive to moderate variations in the initial $\varphi$, since once the drive sufficiently tilts the free-energy landscape, the switching outcome is determined by the resonant excitation and stochastic forces. Both the order-parameter field $\varphi$ and the excitonic envelopes $\psi_j$ are treated as \emph{dimensionless} variables and every numerical quantity that follows is expressed in terms of a characteristic microscopic time chosen as the relaxation time of the order parameter ($\tau_{\varphi}$).
For the generic demonstration here, we adopt ultrafast order-parameter relaxation, fast optical dephasing compatible with low-temperature exciton linewidths, and a comparatively slower thermal relaxation; these choices lie within commonly reported ranges and are not material-specific~\cite{beaurepaire1996ultrafast,giannetti2016ultrafast,kirilyuk2010ultrafast}. The free energy scale $F_0$ is chosen according to $\alpha_0 T_0/F_0 = 36$.  With this convention, we adopt the following set of parameters for the simulations shown in Fig.~\ref{fig:2}: the exciton linewidths satisfy $\gamma_{1}\tau_{\varphi}=\gamma_{2}\tau_{\varphi} = 0.07$, the exciton-exciton non-linearity is given by $a \tau_{\varphi}/\hbar = 0.001$, and the thermal relaxation time $\tau_T/\tau_{\varphi} = 1$ (in a material-specific study, e.g. for \ce{CrI3}, one would typically choose $\tau_T \sim 10\,\tau_\varphi$).  The bath temperature is set as $T_0/T_r = 1.2$, ensuring that the system remains below the equilibrium phase-transition temperature during the entire protocol.  For the shape-selective factor expressed in Eq.~\eqref{eq:psi_dynamics_sel_S}, we set $\Delta\varphi = 0.4$, while the pump strength is quoted in the natural units $P_0 \tau_{\varphi}/\hbar$.  The thermal coefficients are $b_1 \tau_{\varphi}/T_0 = b_2 \tau_{\varphi}/T_0 = 3\times10^{-4}$ and $a_T \tau_{\varphi}/T_0 = 3\times10^{-4}$[$P_0^{-2}$].  In the dimensionless Landau functional [Eq.~\eqref{eq:Landau}], the left coefficients are chosen as $\gamma_0/F_0 = 0.5$, and $\beta/F_0 = 12$.  Finally, the white-noise term is normalized so that $\sqrt{2k_B T_0 \tau_{\varphi} /(F_0 \tau_\eta)} = 0.2$.

Let us start by considering the system's temperature $T<T_0$. Then, the coefficient of the quadratic term in the free energy stays negative $\alpha(T)<0$, placing the system in one of two symmetric local minima in the absence of external pumping, where the cubic term coefficient $\gamma = 0$ and there are no excitons. The symmetric gray solid curve in Fig.~\ref{fig:2}(f) depicts this initial free-energy profile. These minima remain energetically equivalent unless an imbalance arises between the populations of the two exciton phases, making one minimum more favorable. It is important to note, however, that this imbalance alone does not automatically trigger a transition; thermal noise and random forces must still overcome the barrier separating the minima. Here, we should clarify this point. For a continuous resonant pump, the excitonic amplitudes grow until the gain is balanced by losses.  In the long-time limit \(t\gg\gamma_j^{-1}\), Eq.~\eqref{eq:psi_dynamics} gives the stationary population $|\psi_j|_{\mathrm{st}}^{2}\simeq
|P|^{2}/(\gamma_j^{2}+(\omega-\omega_j)^{2})$. Substituting $|\psi_j|_{\mathrm{st}}^{2}$ into $\gamma$ lowers the free-energy minimum of the driven phase and reduces the barrier height $\Delta F$ separating the two minima.  
Because the system's temperature is below $T_0$ as we mentioned, the thermal activation is insufficient: the order parameter $\varphi$ can cross the barrier only through the stochastic force $\eta(t)$. Note that $T_0$ remains fixed as the equilibrium transition temperature; reducing $\Delta F$ via pumping does not alter $T_0$, so we still have $T < T_0$. The instantaneous variance of that force is  
$\sigma_{\eta}^{2}=2 k_B T(t)/ F_0$, so that switching between phases becomes likely when $\sigma_{\eta}\gtrsim \Delta F/F_0$. Thus, a strong pump enhances the steady-state exciton density, softening the barrier while a large noise amplitude (higher \(k_B T\) or $\tau_{\varphi}$) provides the kicks that push $\varphi$ over the residual barrier.  Deterministic switching under continuous illumination therefore emerges from the cooperation between the barrier softening by the stationary exciton reservoir and sufficient Langevin events that complete the transition. In other words, 
the intense pump pulses, tuned to a specific excitonic mode, can destabilize one minimum so that random fluctuations easily push the order parameter into the deeper well. 

To exemplify the mechanism explained above, we use the following form of pump: $P(t)=P_0 \exp{(-{\rm i}\omega_i t)}(\tanh[(t -t_0)/\tau_p]+1)/2$, as shown in Fig.~\ref{fig:2}(d), where we addressed the pump activation function with $\tau_p=0.5\tau_{\varphi}$. When the pump is sufficiently strong, the less favorable minimum becomes so shallow that even moderate noise quickly drives the system toward the global minimum, as illustrated in Fig.~\ref{fig:2}(a) and (f). As discussed earlier, after the exciton population saturation [Fig.~\ref{fig:2}(b) and (c)], one of the minima essentially turns into an inflection point, while the other becomes dominant [panel~\ref{fig:2}(f)]. Consequently, the system ends up in a newly ordered state different from its initial one, confirming the light-induced switching mechanism in the low-temperature regime. This is further evidenced by the temperature remaining below the critical value $T_0$ throughout the process, see Fig.~\ref{fig:2}(e).

\begin{figure}[t!]
\centering
\includegraphics[width=1\linewidth]{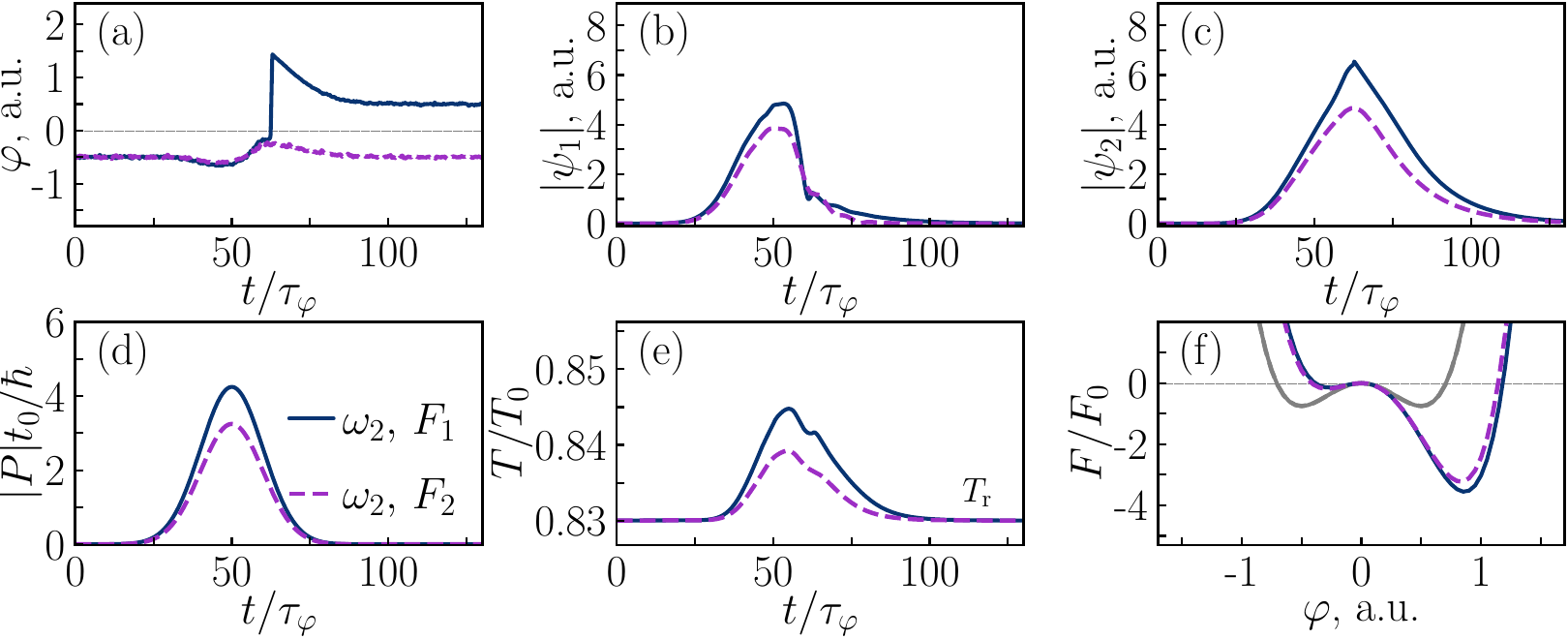}
\caption{Example of single-pulse, fluence-dependent switching. Two identical Gaussian pulses, both resonant with the $\omega_{2}$ exciton, are applied but with different peak amplitudes producing fluences $F_{1}$ (solid blue) and $F_{2}\!<\!F_{1}$ (dashed violet).  
(a)~Order-parameter trajectory $\varphi(t)$.  
The stronger pulse $F_{1}$ transiently suppresses the energy barrier and, helped by Langevin kicks, drives $\varphi$ across the saddle into the opposite minimum, achieving deterministic phase reversal.  
The weaker pulse $F_{2}$ lowers the barrier by a smaller amount; thermal noise then fails to complete the crossing and the system relaxes back to its initial state.  
(b,c)~Populations of the two excitonic modes. Because both pumps are tuned to $\omega_{2}$, the $\psi_{2}$ density (c) demonstrates more intense growth whereas $\psi_{1}$ (b) remains low, evidencing the phase-selective absorption enforced by $S_{j}(\varphi)$.  
(d)~Temporal profiles of the two Gaussian pumps with identical carrier frequency, but different amplitudes.  
(e)~Transient lattice temperature.  Even for $F_{1}$ the rise is modest ($T_{\max}/T_{0}\!\approx\!0.85$), confirming that switching is non-thermal and governed by the exciton-induced cubic term.  
(f)~Landau free-energy density $F(\varphi)$ at $t=0$ (grey, identical for both runs) and at the pulse peak $t=t_{0}$.  
For $F_{1}$ (blue) the cubic contribution tilts the landscape enough to eliminate the left minimum, leaving the right-hand well as the global minimum; for $F_{2}$ (violet) the tilt is insufficient and the system ultimately returns to its original phase. These panels together illustrate the fluence threshold for single-shot, resonant all-optical switching between phases captured by the model.
}
\label{fig:gaussian_pulse}
\end{figure}

In addition to the case of continuous pump, we now consider a protocol in which the external optical pump is modulated by a Gaussian envelope. The system parameters are kept the same. The corresponding system behavior is demonstrated in Fig.~\ref{fig:gaussian_pulse}. In this configuration, the pump is given by $P(t) = P_0\exp[-(t-t_0)^2/2\delta_t^2\ -i\omega t]$ as depicted in Fig.~\ref{fig:gaussian_pulse}(d) , where $P_0$ denotes the peak amplitude and $\delta_t$ defines the the pulse width, with $\delta_t=10\tau_{\varphi}$. The pump fluence is varied by adjusting $P_0$ while keeping $\sigma$ and frequency $\omega$ fixed. As expected, our numerical simulations reveal that for a sufficiently high fluence ($F_1$) the system is driven from its initial phase to an alternate phase, whereas for lower fluences ($F_2$) the system remains in its initial state, see Fig.~\ref{fig:gaussian_pulse}(a).

A key difference from the continuous-wave scenario explored in Fig.~\ref{fig:2} to the Gaussian pulse, is that the latter supplies energy only within a narrow temporal window, so the excitonic reservoir never reaches the stationary value as previously discussed.  Instead, the excitonic populations $|\psi_{1,2}(t)|^{2}$ follow the pump intensity quasi-adiabatically [Figs.~\ref{fig:gaussian_pulse} (b-c)]: they rise rapidly during the leading edge of the pulse, peak near $t\!\simeq\!t_{0}$, and then decay on the intrinsic lifetime $\gamma_{j}^{-1}\!\ll\!\delta_{t}$.  Consequently, both the cubic coefficient $\gamma\rightarrow\gamma(t)=\gamma_{0}\bigl(|\psi_{1}|^{2}-|\psi_{2}|^{2}\bigr)$ and the barrier height $\Delta F \rightarrow\Delta F(t)$, thus changing dynamically.  For the lower-fluence case ($F_{2}$, dashed curve in Fig.~\ref{fig:gaussian_pulse}), the maximum reduction of $\Delta F$ is insufficient to satisfy the noise criterion $\sigma_{\eta}\!\gtrsim\!\Delta F/F_{0}$ at any instant, so $\varphi(t)$ merely oscillates around its original minimum and the system relaxes back to the initial phase once the pulse passes.  

Only when the pulse fluence is increased ($F_{1}$, blue solid curve in Fig.~\ref{fig:gaussian_pulse}), the barrier is flattened a bit more and the Langevin kick occurring close to the pulse maximum is enough to push $\varphi$ over the barrier, after which deterministic relaxation carries the system into the opposite potential well minimum.  The transient temperature increases but still stays below $T_{0}$  as shown in Fig.~\ref{fig:gaussian_pulse}(e), confirming that the phase transition remains non-thermal and is controlled by the interplay between pulse-limited increasing of excitonic population and stochastic activation.  These simulations can therefore reproduce e.g. the experimentally observed fluence threshold for single-shot magnetization reversal in \ce{CrI3} (see next section) and related systems. 

In the context of Gaussian pumps as explored in Fig.~\ref{fig:gaussian_pulse}, it is worth highlighting the importance of matching the pulse duration $\delta_t$ to the exciton lifetime $\gamma_{j}^{-1}$. If $\delta_{t}\ll\gamma_{j}^{-1}$, the excitonic reservoir has no time to develop, and the switching mechanism between phases is suppressed, whereas excessively long pulses approach the continuous-wave limit discussed in Fig.~\ref{fig:2}.

Finally, let us comment qualitatively on the sharpness of the threshold fluence. Allowing for all necessary approximations, we can say that in the presence of Langevin noise, single-shot reversal is intrinsically probabilistic within a narrow fluence window. In a time-dependent Kramers picture~\cite{KRAMERS1940284}, the instantaneous escape rate from the metastable well is
$k(t)\simeq k_0(t)\exp[-\Delta F(t;\mathcal{F})/(k_B T(t))]$, where $\Delta F(t;\mathcal{F})$ is the pulse-induced barrier suppression, $k_0$ is responsible for normalization, and $T(t)$ sets the noise strength through~\eqref{eqn:noise}. Treating barrier crossings as rare and memoryless, the survival probability obeys $\dot S(t)=-k(t)S(t)$, hence
\begin{align}
p(\mathcal{F})=1-\exp\!\Big[-\int k(t;\mathcal{F})\,dt\Big].
\end{align}
Accordingly, $p(\mathcal{F})$ rises from $0$ to $1$ over a width that increases with the noise amplitude and shrinks when the pulse suppresses the barrier more strongly. 
In our simulations this $(F_1,F_2)$ window is so narrow that extensive averaging over noise realizations was unnecessary.

\subsection{Specific LIPT mechanism: All-optical magnetization switching in \ce{CrI3}}\label{sec:spec_mech}  

All-optical magnetization switching (AOS) refers to the reversal of a material magnetization using only ultrafast laser pulses, without any applied magnetic field. Over the last years, a variety of theoretical models have been developed to explain how a short laser excitation can induce magnetic switching in different materials~\cite{stanciu2007all,radu2011,ostler2012,elhadri2016,PhysRevB.104.L020412,PhysRevB.108.094421}. These models generally emphasize two aspects: the transfer of angular momentum from the light to the material lattice magnetization (or within the spin system), and the ultrafast modification of magnetic order parameters (for example, through heating). 

Among the systems where AOS was reported experimentally, the case of chromium tri-iodide (\ce{CrI3}) is of particular interest within our model, as this material combines magnetic ordering with robust excitonic responses, and thus an exciton-mediated mechanism of the optical switching is naturally expected~\cite{PhysRevB.104.L020412}. \ce{CrI3} is a layered ferromagnetic semiconductor whose monolayers exhibit out-of-plane ferromagnetism below about 45~K. Unlike metallic magnets, where magnetization manipulation often relies on broadband absorption and rapid heating, \ce{CrI3} supports discrete excitonic transitions that strongly couple to corresponding magnetic degrees of freedom. According to the microscopic theory, a resonant optical excitation of a particular exciton, capable of carrying angular momentum, can transfer spin angular momentum to the local magnetic moments, reversing the ferromagnetic order with minimal heating. This mechanism requires that the photon energy be tuned to the corresponding electronic transition, making resonant spin-orbit interactions a promising route to ultrafast magnetization control in 2D magnets like \ce{CrI3}.

To grasp this mechanism more concretely, consider a situation where the ground state corresponds to an exciton transition of one helicity (e.g., $\sigma^{-}$ polarization direction), while we illuminate the system with a laser pulse of opposite helicity, i.e., $\sigma^{+}$. Owing to the giant intrinsic Zeeman splitting present in \ce{CrI3}, the $\sigma^{+}$ transition is initially far off-resonance, so that the absorption of $\sigma^{+}$ photons is primarily virtual at first. However, these virtual transitions effectively act like a magnetic field favoring magnetization reversal. Crucially, once the magnetization flips, the $\sigma^{+}$ transition becomes truly resonant, allowing the pump to drive the system into the newly switched state. If the incident pulse intensity exceeds a certain threshold, this feedback between virtual absorption, magnetization reversal, and subsequent real absorption leads to deterministic all-optical switching.

Indeed, Zhang \textit{et al}. (2022)~\cite{Zhang2022} experimentally demonstrated all-optical magnetization switching in a monolayer \ce{CrI3} by tuning circularly polarized laser pulses to a specific excitonic resonance, with the pulse helicity dictating the final magnetization direction. Dabrowski \textit{et al}. (2022)~\cite{Dabrowski2022} then showed a deterministic multi-pulse switching in a CrI$_3$/WSe$_2$ heterostructure, as spin-polarized carriers injected from \ce{WSe2} stabilized the reversed magnetic state. In contrast, bare \ce{CrI3} systems demagnetize randomly under a single pulse, highlighting the vital role of engineered interfaces and spin injection. These observations underscore how carefully tuned optical selection rules, exciton energies, and spin-carrier interactions can yield efficient AOS at relatively low laser fluences, offering an appealing path toward LIPTs for ultrafast control of 2D magnetic materials.

\subsubsection{Adapting our model to CrI$_3$: multi-exciton contribution}

To validate our phenomenological model, we partially simulate the experimental protocol reported by Zhang \textit{et al.} (2022)~\cite{Zhang2022}, in which circularly polarised femtosecond pulses of variable photon energy and fluence reverse the magnetization of a few-layer CrI$_3$. In particular, within our theory, we aim to reproduce  Fig.~3 of Ref.~\citenum{Zhang2022}.

The experiment probes a CrI$_3$ flake initially prepared in the magnetic state $M\!\uparrow$.  It turns out that magnetization reversal is observed only when the pump helicity and photon energy fall into one of two disjoint windows: (i) left-hand circular pump ($\sigma_{-}$) with photon energy $\hbar\omega\simeq1.7\!-\!2.1\;\text{eV}$; or (ii) right-hand  circular pump ($\sigma_{+}$) with $\hbar\omega\gtrsim2.1\;\text{eV}$.  Such selectivity reflects the exceptionally rich excitonic spectrum of CrI$_3$, where both Wannier- and Frenkel-type excitons possessing large oscillator strengths populate the interval 1.6-2.5 eV; each valley couples to a distinct circular polarization, and the high density of states yields several overlapping resonances with markedly different Zeeman shifts~\cite{PhysRevB.108.L161402}.

To reproduce this magnetization reversal mechanism, the basic cubic term $\gamma$ in our model given by Eq.~\eqref{eq:gamma} is no longer sufficient. However, a key advantage of our framework lies in its clarity and versatility, which naturally allows us to write the following generalization:
\begin{equation}
\gamma\!=\!\gamma_{0}\!
\left[
\sum_{k\in\mathcal{K}}
  \bigl(|\psi_{\downarrow,k}^{-}|^{2}\!-\!|\psi_{\uparrow,k}^{+}|^{2}\bigr)
\!+\!\!
\sum_{l\in\mathcal{L}}
  \bigl(|\psi_{\downarrow,l}^{+}|^{2}\!-\!|\psi_{\uparrow,l}^{-}|^{2}\bigr)\!
\right],
\label{eqn:cri3_gamma}
\end{equation}
where $\mathcal{K}=\{1,\dots,N_1\}$, $\mathcal{L}=\{N_1+1,\dots,N_2\}$, with $N_1+N_2$ being the total number of energy-resolved excitonic states. Moreover, $\psi^{\pm}_{\uparrow(\downarrow),k(l)}$ denotes an individual excitonic resonances at $k,l$ states, of given spin $\uparrow,\downarrow$ and helicity $\pm$. Retaining every relevant mode allows us to keep realistic linewidths and pump intensities, whereas a two-mode approximation read as
\begin{align}
    \gamma=\gamma_0\sum_{\sigma=\pm}(|\psi_{\downarrow,\sigma}|^2-|\psi_{\uparrow,\sigma}|^2)
\end{align}
would require artificially broad excitonic lines to match the data. 

\begin{figure}[t]
  \centering
  \includegraphics[width=\linewidth]{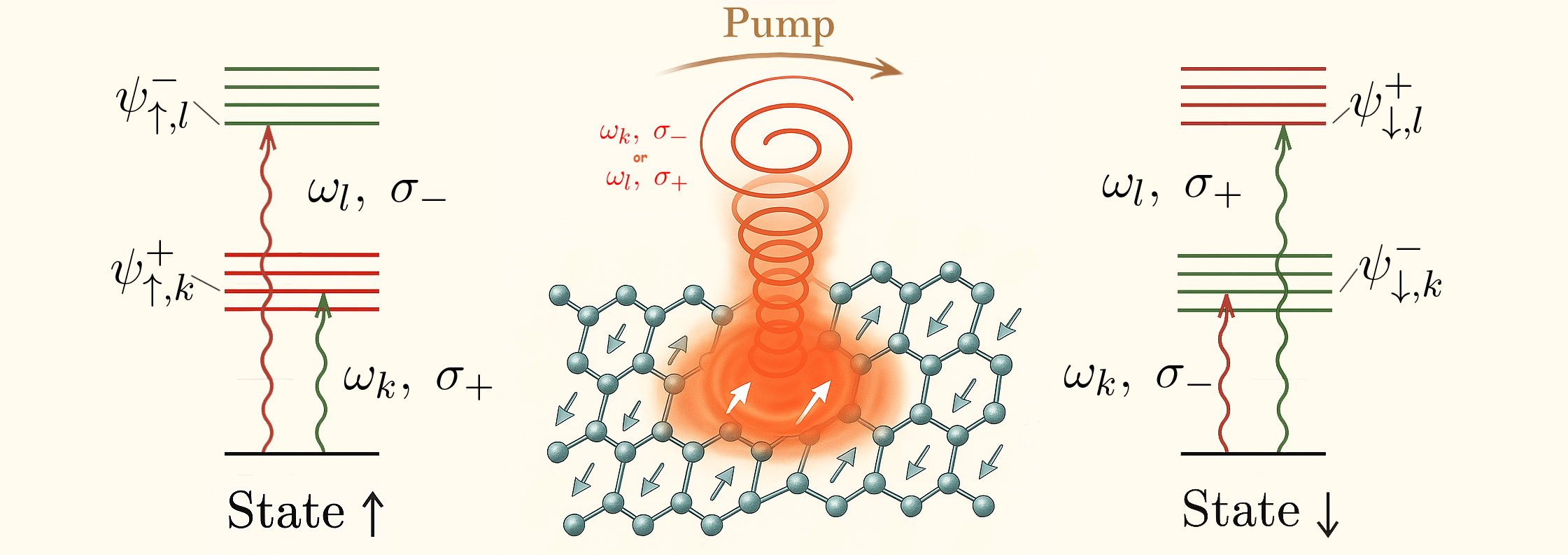}
  \caption{%
Multi-exciton pathway that underlies helicity-frequency-selective magnetization reversal in CrI$_3$. Initially (left panel) the flake resides in the $M\!\uparrow$ ferromagnetic state. Green (red) horizontal lines represent the corresponding spin-down (spin-up) excitons; superscripts $+/-$ label optical helicity $\sigma_{+},\sigma_{-}$ of each ladder. Two ladders are relevant: a lower--energy manifold ($\omega_{k}$) that couples to $\sigma_{+}$, and a higher--energy manifold ($\omega_{l}$) that couples to $\sigma_{-}$ only when the magnetization is~$\uparrow$. A circularly polarised pump (center panel) is tuned into resonance with one of the ladders but carries the \emph{opposite} helicity to the resident magnetic phase. The first photons are therefore absorbed only virtually, yet they seed an exciton population in the \emph{forbidden} ladder of the current phase (excitons with spin-up at $\omega_{l},\sigma_{+}$ or with spin-down at $\omega_{l},\sigma_{-}$). Via the cubic term in Eq.\,\eqref{eqn:cri3_gamma} this unbalanced population lowers the free--energy minimum of the opposite magnetic state and reduces the barrier between wells. When the barrier is sufficiently softened a stochastic fluctuation forces the order parameter across it. After the spin flips to $M\!\downarrow$ (right panel) the same photon energy now addresses an \emph{allowed} ladder; real absorption then rapidly amplifies the exciton reservoir and locks the system into the new minimum. The mechanism is therefore (i) helicity-selective, (ii) threshold-like in fluence, and (iii) non-thermal, being driven by resonant exciton imbalance rather than lattice heating.}
  \label{fig:cri3_scheme}
\end{figure}

Figure~\ref{fig:cri3_scheme} illustrates the AOS mechanism in \ce{CrI3} accounting the multi-exciton contribution of Eq.~\eqref{eqn:cri3_gamma}. The excitonic populations grow during the pulse, unbalancing the two sums in Eq.~\eqref{eqn:cri3_gamma}, lowering the opposing minimum and softening the barrier until a Langevin kick completes the switch even for a slight detuning.

\subsubsection{Reproducing the \ce{CrI3} phase diagram}
For reproducing the experimental finding from Ref.~\citenum{Zhang2022} in CrI$_3$, we discretize each excitonic ladder into forty equidistant modes. The first set spans $1.73-2.12~\textup{eV}$, while the second covers $2.07-2.45~\textup{eV}$. Every mode is assigned the same homogeneous linewidth $\gamma_{j}=15\;\text{meV}$.  The order-parameter relaxation time is fixed at $\tau_{\varphi}=1~\text{ps}$, whereas the thermal relaxation is taken an order of magnitude longer, $\tau_{T}=10~\text{ps}$.  The exciton-exciton non-linearity is kept at $a\tau_{\varphi}/\hbar=10^{-3}$.  The Curie temperature of CrI$_3$ is $T_{0}=45\ \text{K}$ and the bath is set to $T_{r}=5\ \text{K}$.  We retain the overlap width $\Delta\varphi=0.4$ in the shape-selective factor and express the pump amplitude in natural units $P_{0}\tau_{\varphi}/\hbar$.  In the dimensionless Landau functional [Eq.~\eqref{eq:Landau}] the coefficients are chosen as $\alpha_{0} T_{0}/F_{0}=4.5$, $\gamma_{0}/F_{0}=2.5$, and $\beta/F_{0}=12$.  The white-noise amplitude is normalized through $\sqrt{2k_B T_{0}\tau_{\varphi}\bigl/ (F_{0}\tau_{\eta})}=0.2$. Thermal feedback enters via temperature dynamics [Eq.~\eqref{eq:Temperature}] with $b_{j}\tau_{\varphi}/T_{0}=1.1\times10^{-8}$ and $a_{T}\tau_{\varphi}/T_{0}=4\times10^{-9}$[${P_0^{-2}}$]. We employ the same Gaussian pulse as in Fig.~\ref{fig:gaussian_pulse}, so that the drive is pulsed rather than continuous. The reported fluence is expressed as $\mathcal{F}=\kappa P_{0}$, where the prefactor $\kappa$ is the time integral of the normalized envelope and carries all unit conversions.

\begin{figure}[t]
\centering
\includegraphics[width=\linewidth]{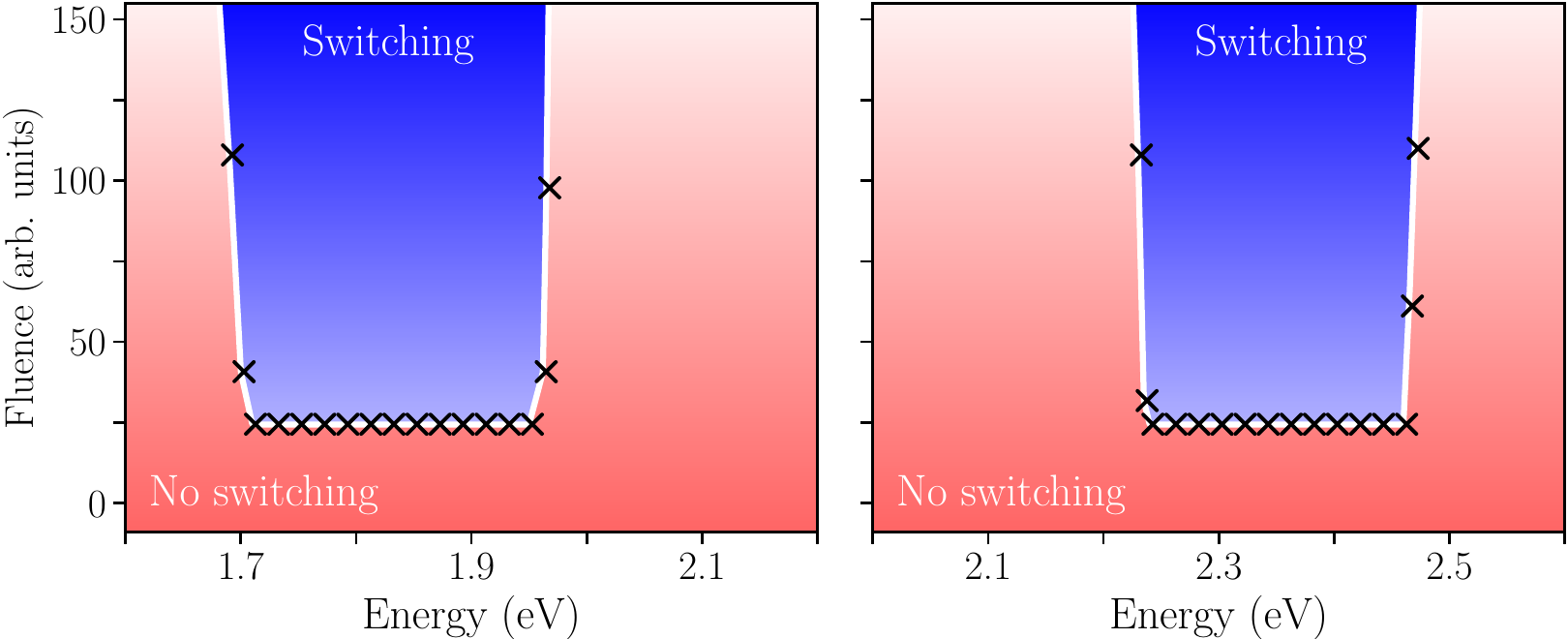}
\caption{%
Phase diagram for single-pulse all-optical switching in CrI$_3$ obtained from the multi-exciton Landau model. Colour encodes the final magnetic state as a function of pump-photon energy (horizontal axis) and fluence (vertical axis): red -- the magnetization relaxes back to its initial $M\!\uparrow$ orientation; blue -- the system flips deterministically to $M\!\downarrow$. Left panel: left-hand circular excitation ($\sigma_{-}$); right panel: right-hand circular excitation ($\sigma_{+}$). Only two narrow \emph{switching pockets} appear, centred at the energy ranges where the pump is resonant with the helicity-opposite exciton ladders sketched in Fig.~\ref{fig:cri3_scheme}.
Inside each pocket reversal occurs only when the pulse fluence exceeds a specific threshold; below this value the cubic exciton term does not fully quench the metastable barrier. Black crosses are explicit simulation points. The topology -- two disjoint, helicity-dependent windows bounded by a sharp fluence threshold -- reproduces the experimental map reported in Ref.~\citenum{Zhang2022}.
}
\label{fig:CrI3_phase}
\end{figure}

The resulting phase maps from our model are shown in Fig.~\ref{fig:CrI3_phase}. They reproduce the three hallmarks of the experiment~\cite{Zhang2022}: a clear switching pocket bounded by a fluence threshold, the sensitivity of that threshold to detuning from the excitonic resonance, and asymmetry between the two opposite circular polarization directions. Although we seeded excitonic ladders across the full windows $1.73\mbox{-}2.12\;\text{eV}$ and $2.07\mbox{-}2.45\;\text{eV}$, the simulated switching pockets are narrower from one specific polarization direction. For $\sigma_{-}$, the left panel of Fig.~\ref{fig:CrI3_phase} shows a reversal behavior starting at $\hbar\omega\!\approx\!1.70\;\text{eV}$ and abruptly terminating near $1.97\;\text{eV}$, even though additional exciton modes still exist above that energy.  Conversely, for $\sigma_{+}$ pump, the model predicts a pocket extending only from $2.23$ to $2.47\;\text{eV}$ (Fig.~\ref{fig:CrI3_phase}, right panel). This happens because the exciton resonances begin to play a role, which corresponds to another stable state of the sample, preventing the barrier from decreasing.

Moreover, in the experimental phase maps from Ref.~\citenum{Zhang2022}, the pocket edges are slightly ragged, whereas our curves remain quite smooth. Introducing a realistic scatter of individual linewidths $\gamma_{j}$, which is unavoidable in any fabricated structure, would imprint comparable irregularities on the simulated boundaries without affecting the overall topology of the phase diagram.

\section{Other systems}

Beyond magnetic monolayers as the \ce{CrI3} explored above, our framework can be applied to soft-lattice semiconductors whose competing phases can be traced back to distinct excitonic flavours rather than to spin order. A prime example is the two-dimensional Ruddlesden-Popper lead-halide perovskites, where time-resolved spectroscopy shows that photo-excitations oscillate between a delocalised free-exciton (FE) polaron and a momentarily self-trapped exciton (STE) that is pinned by local lattice distortions on a sub-picosecond time-scale~\cite{Tao2021Momentarily}. 

Here the two minima of the Landau potential correspond to the FE-polaron and STE geometries, while the order parameter $\varphi$ can be chosen as the lattice distortion amplitude that accompanies self-trapping. Because the FE and STE manifolds possess different optical selection rules and phonon-broadened spectral profiles, resonant pumping at the FE transition (\emph{e.g.}, the high-energy side of the asymmetric photoluminescence band) preferentially populates one excitonic field, say $\psi_{1}$, whereas selective excitation of the STE tail enhances the other, $\psi_{2}$.  The ensuing imbalance $|\psi_{1}|^{2}-|\psi_{2}|^{2}$ enters our cubic term $\gamma\varphi^{3}$ exactly as in Eq.~\eqref{eq:gamma}, tilting the free-energy landscape and steering the crystal between free and self-trapped regimes without requiring any global heating. In this way the model offers a mesoscopic route to interpret the fluence-dependent quenching of the low-energy PL tail and the ultrafast spectral breathing reported for CsPbBr\textsubscript{3} nanoplatelets.

A second, structurally driven, illustration is provided by thin-film MAPbI\textsubscript{3} where tetragonal and cubic nanodomains coexist at room temperature and self-organise into double- and triple-layer superlattices~\cite{Kim2018SelfOrganized}.  
Each crystallographic variant hosts its own set of excitons, so that circularly-polarized or energy-tuned pulses can again tilt the double-well potential by selectively building an exciton reservoir in just one domain type.  Our Landau-exciton scheme therefore explains how light can assist the observed room-temperature migration of phase boundaries, or stabilize metastable superlattice buffers, by lowering the free energy of the optically favored domain through the same exciton-mediated cubic coupling.  

Taken together, these two perovskite case studies underline the generality of our approach: whenever (i) two competing phases are connected to spectrally distinguishable excitonic transitions, and (ii) those excitons can be driven far from equilibrium, the simple set of Eqs.~\eqref{eq:EOM_phi}-\eqref{eq:Temperature} is sufficient to capture light-induced switching between them.  We therefore anticipate immediate applications to other hybrid semiconductors displaying free/STE duality, to halide-perovskite polytypes exhibiting local ferroelastic twins, and to any material where optical pumping can discriminate between coexisting excitonic reservoirs.

Let us also add, that beyond the spatially uniform setting considered here, the framework admits a straightforward inhomogeneous generalization. Adding a gradient term $\kappa|\nabla\varphi|^2$ to the Landau functional and, optionally, exciton diffusion (either at the level of populations, $\partial_t n_j=\dots+D_j\nabla^2 n_j$, or within the complex-field formulation) yields a time-dependent Ginzburg–Landau/phase-field model for $\varphi(\mathbf r,t)$ locally coupled to $S_j(\varphi(\mathbf r,t))$. This connects our approach to domain-wall kinetics, curvature-driven relaxation, and noise-assisted nucleation under resonant optical driving, and provides a natural route for future spatially resolved studies without changing the mechanisms established here.

\section{Conclusion}
 
We developed a phenomenological model based on the Landau theory to describe light-induced phase transitions (LIPTs) specifically triggered by resonantly generated excitons under ultrafast pump pulses. The framework incorporates dynamic coupling between the order parameter, temperature, complex excitonic fields, and stochastic Langevin forces. By applying this model, we successfully reproduced key features of experimentally observed all-optical magnetization switching in a \ce{CrI3} layer, including the sharp fluence threshold, strong dependence on exciton-photon detuning, and asymmetric response to pumps of opposite $\sigma_-$ and $\sigma_+$ helicities. In principle, the model can be extended to any class of LIPTs characterized by a well-defined order parameter and mediated by resonantly excited excitonic states, offering a general framework to both explain existing phenomena and guide the exploration of new systems supporting optically induced phase transition mechanisms.

\begin{acknowledgments}
The work of A.K. is supported by the Icelandic Research Fund (Ranns\'oknasj\'o{\dh}ur, Grant No.~2410550). L.S.R. acknowledges the Icelandic Research Fund (Rannís), grant No. 239552-051. H.S. acknowledges the project No. 2022/45/P/ST3/00467 co-funded by the Polish National Science Centre and the European Union Framework Programme for Research and Innovation Horizon 2020 under the Marie Skłodowska-Curie grant agreement No. 945339.
\end{acknowledgments}

\section*{Author Declarations}

\subsection*{Conflict of Interest}
The authors have no conflicts to disclose.

\subsection*{Author Contributions}
\textbf{A. Kudlis:} Conceptualization (lead); Formal analysis (lead); Investigation (equal); Methodology (lead);
Project administration (lead); Supervision (lead);
Visualization (lead); Writing – original draft (lead);
Writing – review \& editing (equal); Funding acquisition (lead).
\textbf{L.S. Ricco:} Conceptualization (supporting); Formal analysis (supporting); Investigation (equal); Methodology (supporting);
Project administration (supporting);
Visualization (supporting); Writing – original draft (lead);
Writing – review \& editing (equal).
\textbf{H. Sigur{\dh}sson:} Conceptualization (supporting); Formal analysis (supporting); Investigation (supporting);
Project administration (supporting); Writing – original draft (supporting); Writing – review \& editing (equal). Fund acquisition (lead).
\textbf{I.A. Shelykh:} Conceptualization (lead); Formal analysis (lead); Investigation (supporting); Methodology (lead);
Project administration (lead); Supervision (lead);
Visualization (supporting); Writing – original draft (lead);
Writing – review \& editing (equal).

\section*{Data Availability}
The data that support the findings of this study are available from the corresponding author upon reasonable request.

\section*{References}
\bibliography{aip_lit}

\end{document}